\documentclass[iop,apjl,twocolumn]{emulatedapj}
\usepackage{natbib}
\usepackage{graphicx}
\usepackage{rotating}
\usepackage{times}
\usepackage{amsmath}
\usepackage{txfonts}
\usepackage{multirow}

\newcommand{\msun}{{\,\rm M}_{\odot}}

\newcommand{\nht}{\ifmmode {{\rm NH}_3} \else {NH{\bas 3}} \fi}
\newcommand{\tco}{\ifmmode {^{13}{\rm CO}} \else {$^{13}{\rm CO}$}\fi}
\newcommand{\dco}{\ifmmode {^{12}{\rm CO}} \else {$^{12}{\rm CO}$}\fi}
\newcommand{\cdo}{\ifmmode {{\rm C}^{18}{\rm O}} \else {${\rm C}^{18}{\rm O}$}\fi}

\newcommand{\htco}{\ifmmode {{\rm H}^{13}{\rm CO}^{+} } \else {${\rm H}^{13}
{\rm CO}^{+}$ }\fi}
\newcommand{\hco}{\ifmmode {{\rm H}^{12}{\rm CO}^{+} } \else {${\rm H}^{12}
{\rm CO}^{+}$ }\fi}
\newcommand{\juz}{\ifmmode {{\rm J}=1\rightarrow 0} \else
{J=1$\rightarrow$0}\fi}
\newcommand{\jdu}{\ifmmode {{\rm J}=2\rightarrow 1} \else
{J=2$\rightarrow$1}\fi}
\newcommand{\jtd}{\ifmmode {{\rm J}=3\rightarrow 2} \else
{J=3$\rightarrow$2} \fi}
\newcommand{\jcq}{\ifmmode {{\rm J}=5\!\rightarrow\!4} \else
{${\rm J}=5\!\rightarrow\!4$} \fi}
\newcommand{\as}{\ifmmode {^{\scriptscriptstyle\prime\prime}}
        \else $^{\scriptscriptstyle\prime\prime}$\fi}
\newcommand{\am}{\ifmmode {^{\scriptscriptstyle\prime}}
        \else $^{\scriptscriptstyle\prime}$\fi}
\newcommand{\hh}{\ifmmode {{\rm H}_2} \else {H$_2$} \fi}
\renewcommand{\hco}{\ifmmode {{\rm HCO}^+} \else {HCO$^+$} \fi}
\newcommand{\hhco}{\ifmmode {{\rm H}_2{\rm CO}} \else {H$_2$CO} \fi}
\newcommand{\ddco}{\ifmmode {{\rm D}_2{\rm CO}} \else {D$_2$CO} \fi}
\newcommand{\chhdoh}{\ifmmode {{\rm CH}_2{\rm DOH}^+} \else {CH$_2$DOH} \fi}
\newcommand{\chhhod}{\ifmmode {{\rm CH}_3{\rm OD}^+} \else {CH$_3$OD} \fi}
\newcommand{\chhhoh}{\ifmmode {{\rm CH}_3{\rm OH}^+} \else {CH$_3$OH} \fi}
\newcommand{\tchhhoh}{\ifmmode {^{13}{\rm CH}_3{\rm OH}^+} \else {$^{13}$CH$_3$OH} \fi}
\newcommand{\dcop}{\ifmmode {{\rm DCO}^+} \else {DCO$^+$} \fi}
\newcommand{\cchh}{\ifmmode {{\rm C}_2{\rm H}_2} \else {C$_2$H$_2$} \fi}

\newcommand{\hcccn}{\ifmmode {{\rm HC}_3{\rm N}} \else {HC$_3$N}\fi}

\newcommand{\hhd}{H$_2$D$^+$}

\newcommand{\cp}{C$^+$}

\begin{document}

\title{{CID: Chemistry In Disks \\
VII. First detection of HC$_3$N in protoplanetary disks}
\thanks{Based on observations carried out with the IRAM 30m radiotelescope.
IRAM is supported by INSU/CNRS (France), MPG (Germany) and IGN (Spain). 
}}

\shortauthors{Chapillon et al 2012}
\shorttitle{First detection of HC$_3$N}

\author{Edwige Chapillon}
\affil{Institute of Astronomy and Astrophysics, Academia Sinica, P.O. Box 23-141, Taipei 106, Taiwan, ROC}
\email{chapillon@asiaa.sinica.edu.tw}
\author{Anne Dutrey, St\'ephane Guilloteau}
\affil{Univ. Bordeaux, LAB, UMR 5804, 2 rue de l'observatoire, F-33270, Floirac, France}
\affil{CNRS, LAB, UMR 5804, F-33270 Floirac, France}
\email{dutrey@obs.u-bordeaux1.fr, guilloteau@obs.u-bordeaux1.fr}
\author{Vincent Pi\'etu}
\affil{IRAM, 300 rue de la piscine, F-38406 Saint Martin d'H\`eres, France}
\email{pietu@iram.fr}
\author{Valentine Wakelam, Franck Hersant}
\affil{Univ. Bordeaux, LAB, UMR 5804, 2 rue de l'observatoire, F-33270, Floirac, France}
\affil{CNRS, LAB, UMR 5804, F-33270 Floirac, France}
\email{wakelam@obs.u-bordeaux1.fr, hersant@obs.u-bordeaux1.fr}
\author{ Fr\'ederic Gueth}
\affil{IRAM, 300 rue de la piscine, F-38406 Saint Martin d'H\`eres, France}
\email{gueth@iram.fr}
\author{Thomas Henning, Ralf Launhardt}
\affil{Max-Planck-Institut f\"ur Astronomie, K\"onigstuhl 17, D-69117
Heidelberg, Germany}
\email{henning@mpia.de, rl@mpia.de}
\author{Katharina Schreyer}
\affil{Astrophysikalisches Institut und Universitäts-Sternwarte, Schillergässchen 2-3, 07745 Jena, Germany}
\email{k.schreyer@uni-jena.de}
\and
\author{Dmitry Semenov}
\affil{Max-Planck-Institut f\"ur Astronomie, K\"onigstuhl 17, D-69117 Heidelberg, Germany}
\email{semenov@mpia.de}

\date{Received **-***-****, Accepted **-***-****}

  \begin{abstract}

Molecular line emission from protoplanetary disks is a powerful tool to constrain their physical and chemical structure. Nevertheless, only a few molecules have been detected in disks so far.
We take advantage of the enhanced capabilities of the IRAM 30m telescope  by using the new broad band correlator (FTS) to  search for so far undetected molecules in the protoplanetary  disks surrounding the TTauri stars DM Tau, GO Tau, LkCa15 and the Herbig Ae star MWC\,480.
 We report the first detection of HC$_3$N at 5$\sigma$ in the GO Tau and MWC 480 disks with the IRAM 30-m, and in the LkCa 15 disk (5 $\sigma$), using the IRAM array, with derived column densities of the order of $10^{12}$cm$^{-2}$.
We also obtain stringent upper limits on
CCS (N $< 1.5 \times\ 10^{12} \textrm{cm}^{-3}$). We discuss the observational results by comparing them to column densities derived from existing chemical disk models (computed using the chemical code Nautilus) and based on
previous nitrogen and sulfur-bearing molecule observations.
The observed column densities of HC$_3$N are typically two orders of magnitude lower than the existing predictions and appear to be lower in the presence of strong UV flux, suggesting that the molecular chemistry is sensitive to the UV penetration through the disk.
The CCS upper limits reinforce our model with low elemental abundance of sulfur derived from other sulfur-bearing molecules (CS, H$_2$S and SO).

\end{abstract}


%----------------------------------
\maketitle{}
%----------------------------------

%-------------------------------------------------------------------
\section{Introduction}

Understanding the formation of planetary systems requires an in-depth study of the initial conditions, i.e. the structures of the protoplanetary disks. 
Several theoretical works have been done \citep[e.g.][and references therein]{Bergin_etal2007}, leading to the current picture of a flared disk consisting of three layers \citep[e.g.][]{vanZadelhoff_etal2001,Bergin_etal2007}.
However, key model parameters such as gas density and temperature remain so far poorly constrained by observations due to the  limited sensitivity and spatial resolution of current instruments.
Observations of molecular lines have proven to be an excellent tool  to study the physical and chemical structure and the dynamics of protoplanetary disks \citep[e.g.][]{Dutrey_etal1997,Kastner_etal1997,Guilloteau_Dutrey_1998,Pietu_etal2007,Qi+etal2008,Semenov_Wiebe2011,Oberg_etal2012}.

Depending on the molecule and transition, the observations trace different physical conditions and, therefore, sample chemically different regions in the disks.
In gas-rich protoplanetary disks
the \cp\ emission likely comes from the ionized upper part of disks \citep{Semenov_etal2004, jonkheid_etal2007, chapillon_etal2010, panic_etal2010, kamp_etal2011, bruderer_etal2012}. Atomic species such as O{\small I} have also been detected by the Herschel satellite but its origin (warm atmosphere, disk wind or jet?) is still debated \citep{Mathews+2010}. CO and its main isotopomers are characteristics of the molecular layers which are located
up to a few scale heights only above the mid-plane \citep{Dartois_etal2003}.
Near the disk plane, the gas is very cold, so that molecules are expected to be depleted on grains. Such regions may be only traced by \hhd\ \citep{chapillon_etal2012, oberg_etal2011-h2d}.
To retrieve useful information and to reconstruct the 3D (physical and chemical) structure of disks, molecular observations need to be compared with chemical models dedicated to protoplanetary disks \citep{Semenov_etal2010, Vasyunin_etal2011}.
Such studies help characterise the dominant processes in the disk leading to planet formation (e.g., grain growth, molecule formation and destruction, gas-to-dust ratio evolution).

So far, in the mm/submm range which characterizes the bulk of the gas disk, only a few molecules have been firmly detected in protoplanetary
disks: CO and its main isotopomers (\tco~and \cdo), HCO$^+$ and H$^{13}$CO$^+$, DCO$^+$, H$_2$CO, H$_2$O, CS, C$_2$H, N$_2$H$^+$, HCN, HNC, CN and DCN \citep{Dutrey_etal1997,vanDishoeck+etal2003,dutrey_etal2007,Qi+etal2008,Henning_etal2010, oberg_etal2011,Hogerheijde_etal2011, Oberg_etal2012}.
These are simple, light (mass number $< 44$) molecules. Heavier or
more complex molecules remain undetected 
 due to their low abundances and the lack of sensitivity of current instruments.

Within the framework of the CID ("Chemistry In Disks") collaboration, we take here advantage of the improved performance of the IRAM-30m telescope to carry out a molecular survey of four well known large (R$_{out}>500$\,AU) protoplanetary disks. The main driver
of our observational project was to search for heavier molecules.
We report in this observational paper the detection of the new molecule \hcccn ,  in the three disks surrounding the T\,Tauri stars GO Tau and  LkCa15 and the Herbig Ae star MWC\,480. We also present the upper limits obtained on CCS after a deep integration.
A more complete chemical analysis will be presented in a forthcoming paper (Wakelam et al. in prep.).

In Section 2, we present the observations and the methods used for the data reduction. Results (derivation of the best-fit model and column densities) are  described in Section 3. In Section 4 the implications of our new results are discussed. Summary and conclusion are presented in Section 5.

\section{Source selection and observations}
\label{sec:obs}

We focus our survey on well-known molecular-rich protoplanetary disks:
 MWC\,480, LkCa15 and DM Tau \citep{Dutrey_etal1997,Guilloteau_Dutrey_1998,Henning_etal2010,oberg_etal2010}, and GO Tau \citep{Schaeffer_etal2009}. MWC\,480 is an Herbig Ae star with a stellar mass of 1.8 $\msun$ \citep{Simon_etal2000}, while  DM Tau and LkCa15 are T\,Tauri stars of mass 0.5 and 1.0 $\msun$, respectively.  Their disk structures are well known \citep{Pietu_etal2007}. A large CO disk ($\sim 900$~AU) around the 0.5 $\msun$ star GO Tau was reported by \citet{Schaeffer_etal2009}. This disk is similar in many aspects to that surrounding DM Tau \citep{Dutrey+etal2011}. Sulfur-bearing molecules have been recently searched for in these four disks
by the CID collaboration \citep[][hereafter CIDV]{Dutrey+etal2011}.

Observations were carried out in August 2011 with the IRAM 30-m telescope under quite good weather conditions using the wobbler switch and, following our standard observing mode \citep{Dutrey_etal1997,Dutrey+etal2011}. We used the new 30-m heterodyne receiver EMIR. All the lines were simultaneously observed using one frequency setup thanks to the capability of the new broad-band spectrometer (FTS). The setup covered the \hcccn\ J=10-9, J=12-11 and J=16-15 rotational lines (at frequencies 90.978, 109.173 and 145.560 GHz)
and the CCS (7,7-6,8), (7,8-6,7) and (11,12-10,11) transitions at 90.686, 93.870 and 144.244\,GHz respectively.
The frequency resolution was 200 kHz (i.e. 0.6 -- 0.4 km.s$^{-1}$). Frequent pointing and focus checks were done on Jupiter, Mars and the quasar 0528+134.
 IK\,Tau/NML Tau served as line calibrator.
 We spent 5 to 12 hours on each source and the T$_\mathrm{sys}$ was 110-130 K at 3 mm and 135-200 K at 2 mm.
The observations were converted in unit of flux density using the IRAM recommendations\footnote{http://www.iram.es/IRAMES/mainWiki/Iram30mEfficiencies}.

We also obtained low spectral resolution (8 km.s$^{-1}$) images of LkCa15 and GO Tau with 3.5'' angular resolution
using the IRAM Plateau de Bure interferometer, covering the frequency band 144.65 - 148.35 GHz.
Observations were performed on 30-Jul and 14-Sep 2011 in the CD configuration.
Standard phase and amplitude calibration were applied, and the flux calibration used MWC 349 as a reference.

\begin{figure*}%[t!]
\begin{center}
\includegraphics[angle=270,width=0.9\textwidth]{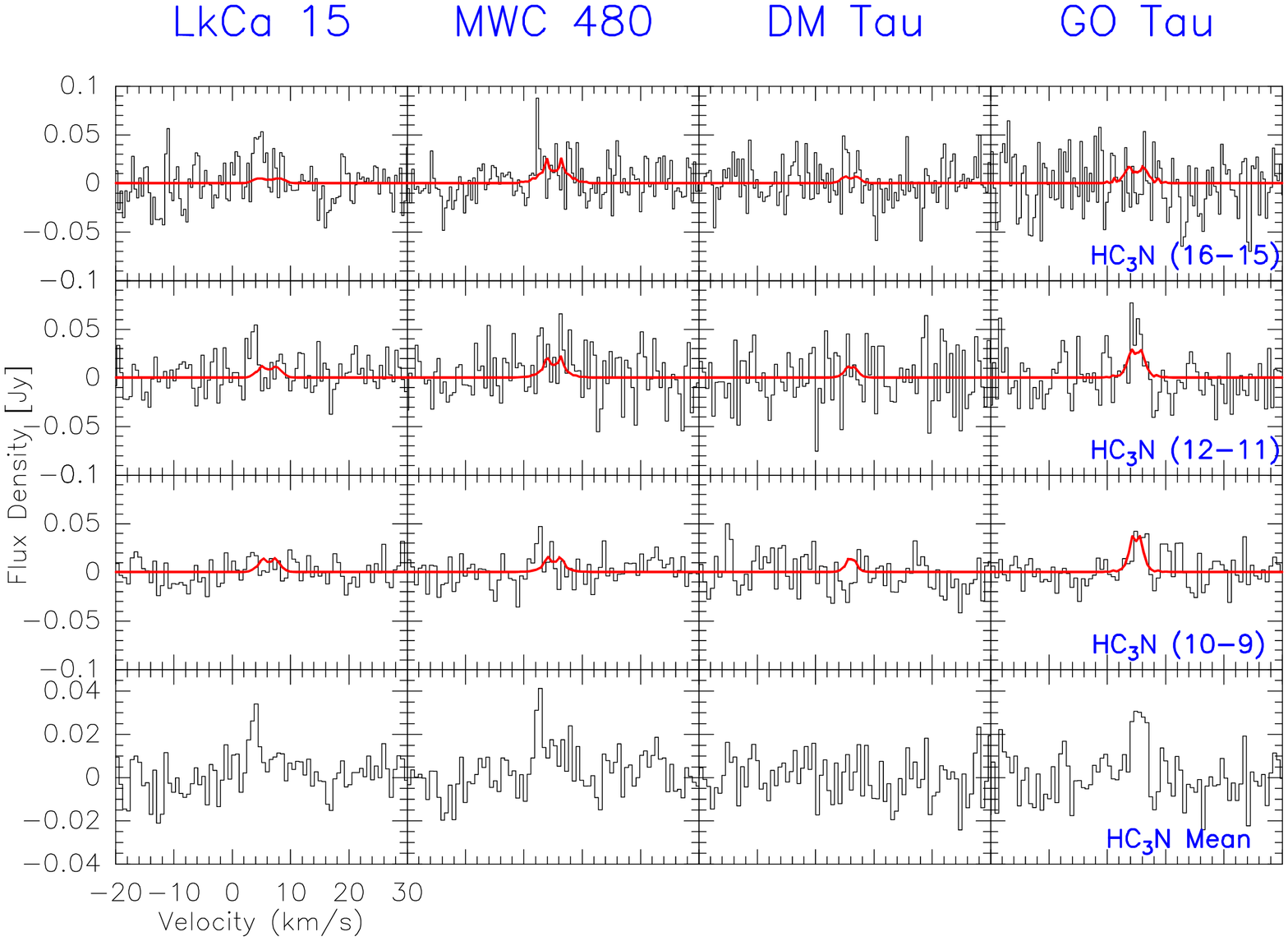}\\
\includegraphics[angle=270,width=0.9\textwidth]{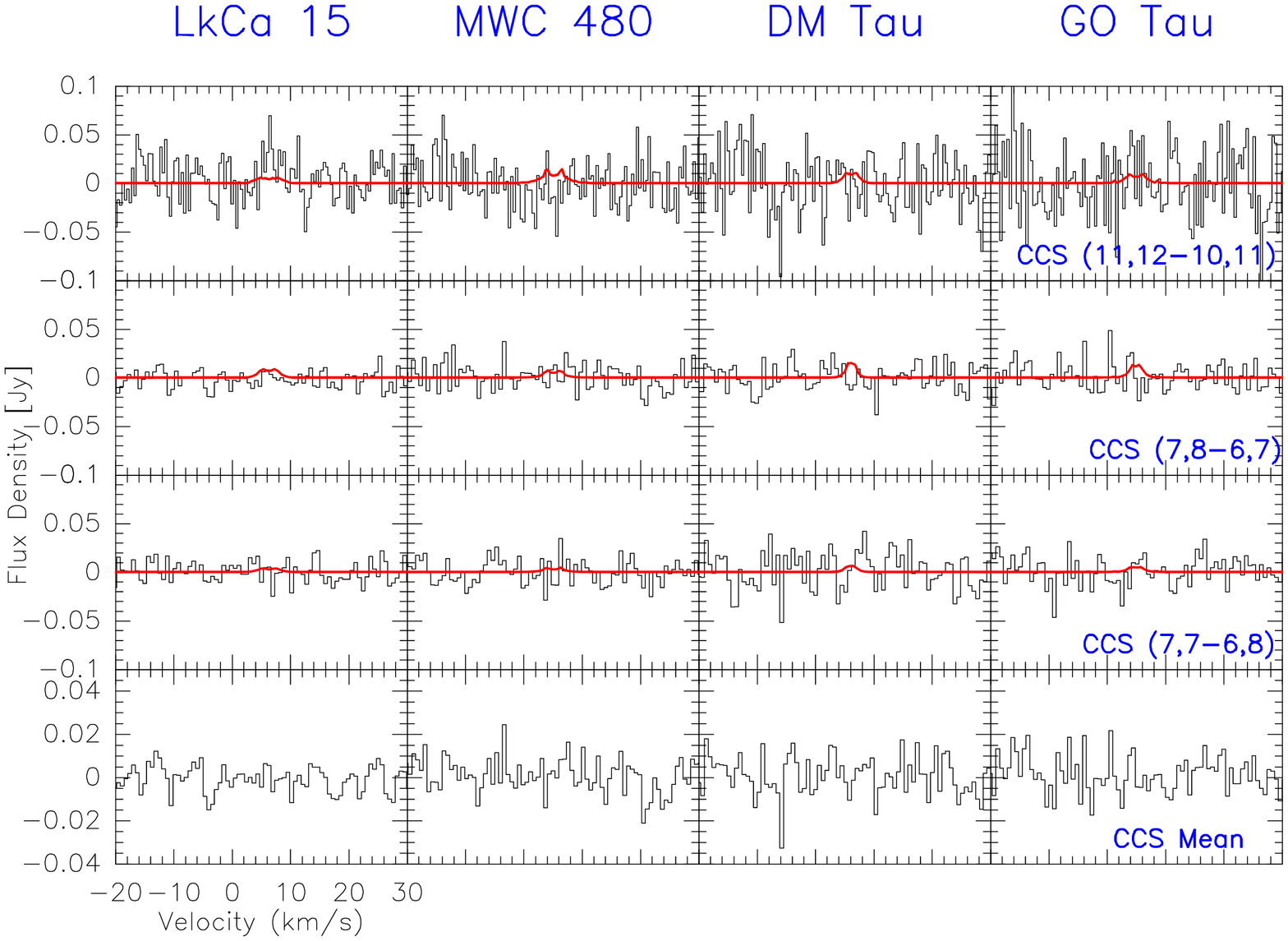}\\
\end{center}
\caption{Observations of \hcccn\ and CCS in the four disks.
Top panel: \hcccn . Bottom panel: CCS.
In both panels, the last line is the average of the three observed transitions.
The best fits or the 3$\sigma$ upper limits of the best model are also superimposed
on the individual lines.}
\label{fig:obs}
\end{figure*}

\begin{figure}%[t!]
%\begin{center}
\includegraphics[angle=0,width=0.4\textwidth]{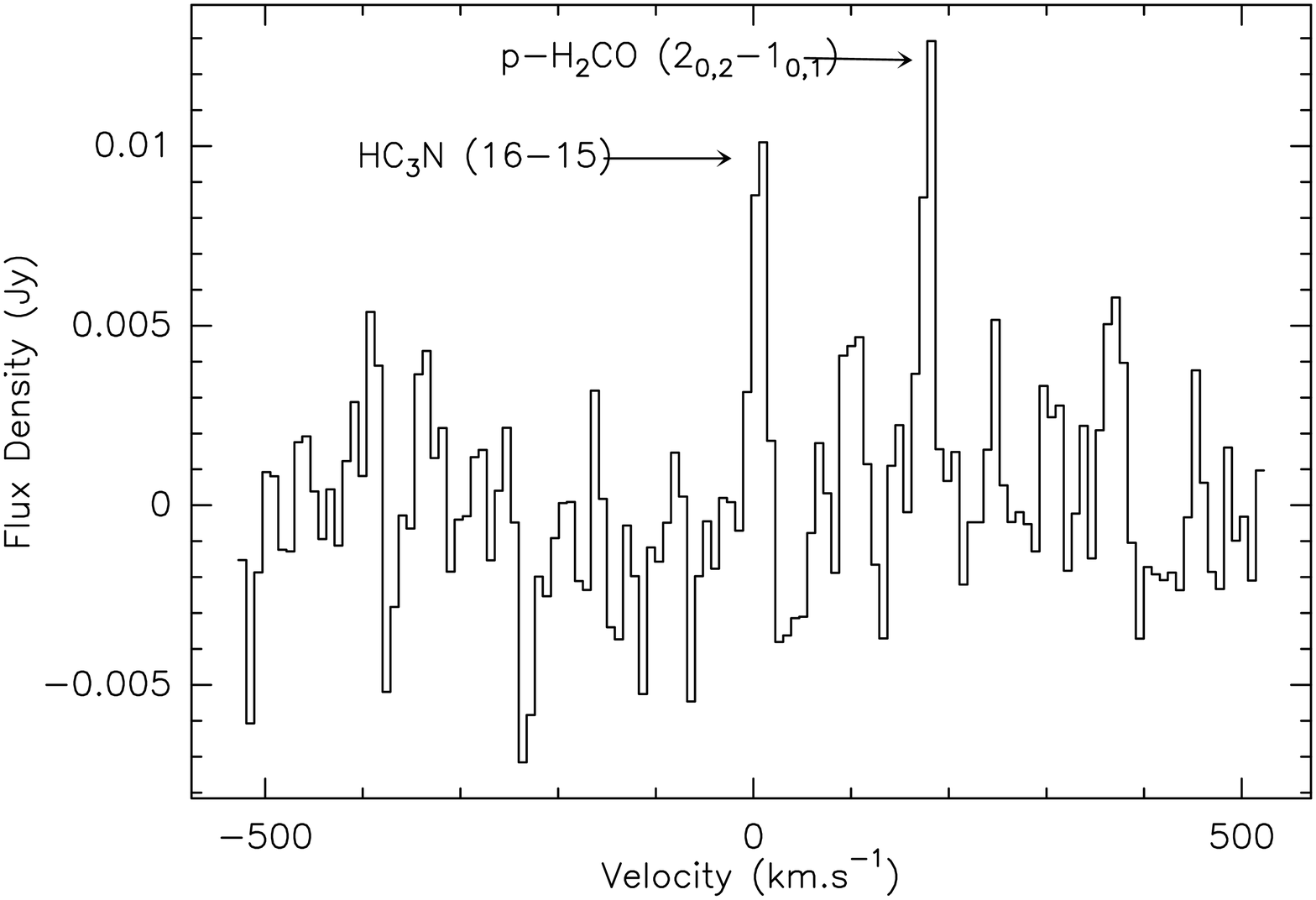}\\
%\end{center}
\caption{ Spectrum of \hcccn\ and H$_2$CO towards LkCa15 obtained at 3.5$''$ resolution with the IRAM array.
The two detected lines are indicated.}
\label{fig:pdbi}
\end{figure}

The data was reduced using the GILDAS\footnote{http://www.iram.fr/IRAMFR/GILDAS/} package.

\section{Results and Column density derivation}
\label{sec:result}

Figure \ref{fig:obs} is a montage showing the 30-m observational results.
We emphasize the sensitivity by summing up all observed transitions of each molecule. The averaged spectra are presented in the bottom panels.
CCS is not detected but we get good upper limits (see Table \ref{tab:fluxes}). The first cyanopolyyne \hcccn\ is convincingly detected in two disks at the level of 5$\sigma$ (GO Tau and MWC\,480) while it is  marginal in LkCa15 (4 $\sigma$) on the integrated spectra and undetected in DM Tau.
 MWC\,480, DM Tau and LkCa15 are located in ``holes'' of their parent molecular clouds in regions
devoid of CO. This is not the case of GO Tau; \citet{Schaeffer_etal2009} reported from PdBI data that the red part
of the CO J=1-0 and J=2-1 lines are hidden by the presence of CO at the same velocity. We carefully checked if confusion can affect the \hcccn\ flux density in GO Tau. From \citet{Schaeffer_etal2009} and $^{13}$CO data obtained simultaneously with the \hcccn\ observations, we estimate which velocity channels could suffer from contamination. While the $^{13}$CO data still present some confusion with the CO ambient cloud, this is not significant for
the \hcccn\ lines. Confusion, if any, would rather absorb the red-shifted part of the spectrum than lead to excess emission,
because the critical densities for substantial excitation of the J=11 to J=16 rotational levels is high. Any
analysis using only the unconfused half of the disk velocity profile would lead to an increase of the \hcccn\ column density estimate.
For GO Tau, only an upper limit of $<0.1$ Jy.km/s can be reported for H$_3$CN (16-15) from the PdBI observations,
compatible with the upper limit given in Table \ref{tab:fluxes}.

For LkCa 15, the blue-shifted part appears somewhat stronger than the red-shifted one.  Spectra of $^{13}$CO taken simultaneously
do not show such a behavior, so this cannot be ascribed to mis-pointing. Note that this 
effect is however at the Signal to Noise limit.
Figure \ref{fig:pdbi} shows the spectrum towards LkCa15 at 3.5$''$ spatial resolution obtained with the PdBI,
after removal of the 20 mJy/beam continuum emission given by the wideband correlator.
The velocity scale refers to  the rest frequency of
the J=16-15 transition of HC$_3$N at 145.560938 GHz. The line is clearly detected ($S/N > 5$), with a total
line flux around 0.15 Jy.km/s, and is about a factor 1.5 weaker than the $2_{0,2}-1_{0,1}$ transition of para
H$_2$CO at 145.602953 GHz also visible in this spectrum. These observations clearly show that the emission
detected with the 30-m is indeed coming from the circumstellar disk.

\begin{table*}
%{\footnotesize
\caption{IRAM 30-m integrated line fluxes for \hcccn\ lines and 3$\sigma$ upper limits for CCS (Jy.km.s$^{-1}$)}
\centering
\begin{tabular}{lcccc}
\tableline \tableline
 Sources          & LkCa\,15 & MWC\,480 & DM\,Tau & GO\,Tau\\
\hcccn\ (16-15) & $0.10 \pm 0.03$ & $0.090 \pm 0.029$ & $\leq 0.06$ & $\leq$0.08\\
\hcccn\ (12-11) &$\leq 0.09 $ ($0.034 \pm 0.031$) & $0.132 \pm 0.040$ & $\leq 0.09$ & $0.102 \pm 0.024$ \\
\hcccn\ (10-9)  &$\leq 0.06 $ ($0.025 \pm 0.022$) & $0.060 \pm 0.026$ & $\leq 0.06$ & $0.067 \pm 0.018$ \\
CCS (11,12-10.11) & $\leq 0.36$ & $\leq 0.11$ & $\leq 0.08$ & $\leq 0.09$\\
CCS (7,8-6,7) & $\leq 0.05$ & $\leq 0.08$ & $\leq 0.05$ & $\leq 0.05 $ \\
CCS (7,7-6,8) & $\leq 0.06 $ & $\leq 0.08$ & $\leq 0.06$ & $\leq 0.05$ \\
\tableline
\end{tabular}
\tablecomments{Integrated fluxes are derived from gaussian fits with fixed central velocities ($V_{sys}$ given in
Tab.\ref{tab:diskfit}) and line-widths corresponding to Keplerian motions with $\Delta V = 1.4$ km.s$^{-1}$  for GO Tau and DM Tau,
$\Delta V = 4.1$ km.s$^{-1}$ for MWC\,480 and $\Delta V = 3.3$ km.s$^{-1}$ for LkCa15.}
\label{tab:fluxes}
%}
\end{table*}

We retrieve the molecular column densities from the data using DiskFit \citep[a dedicated radiative transfer code for protoplanetary disks, see][]{Pietu_etal2007} and following the standard method described in detail in CIDV where the physical disks models are parameterized by power
laws.  The disk geometry and kinematics are consistently incorporated in the model. In particular, we properly take into account the velocity gradient due to Keplerian rotation. At a given velocity, only one well defined portion of the disk emits and hence contributes to the observed column density. All parameters were fixed (see Table \ref{tab:diskfit}) except the molecular surface densities (see Section 2.2 of CIDV for more information). Since we have observed three lines of the same molecule, both for CCS and \hcccn , the best disk model
is obtained by fitting simultaneously these three lines, as we proceeded in \citet{Pietu_etal2007} for CO. Note that the best model is obtained
using a natural weighting (system temperature only) of each line. As in the analysis of the sulfur data (CIDV),
we assume  for the surface density power law versus radius with an exponent $p=1.5$ ($\Sigma(r) = \Sigma_{300} \times (r/{300\mathrm{AU}})^{-1.5}$ and for the temperature $T(r) = T_{100} (r / 100\mathrm{AU})^{-q}$ with $q =0.4$. As a radius of reference, we use 300 AU because it is representative of the area where the unresolved and beam diluted emission arises. A smaller value would be unrealistic because
of the beam dilution effect, a larger value would correspond to a disk area where lines would be too optically thin. For GO Tau,
the source with the highest S/N ratio, we also tried to fit simultaneously $p$ with $\Sigma_{300}$ and found $p = 1.4 \pm 0.3$.
Table \ref{tab:diskfit}, adapted from CIDV, summarizes the physical disk properties used to retrieve the best fit disk model.
The derived surface densities at 300 AU are
presented in Table \ref{tab:results-ratio}.
The line profiles obtained with the best disk models are overplotted in Fig.\ref{fig:obs} on the observations. The best fits do
not follow the individual spectra because it is the result of a simultaneous fit of the three lines giving more weight to the best
S/N ratio spectrum.

\begin{table}
\caption{Disk physical parameters used to derive the best fit models}
\centering
\begin{tabular}{lcccc}
\tableline
\tableline
Source                 &  LkCa~15 & GO~Tau& DM~Tau & MWC~480   \\
\tableline
inclination ($^{o}$)   & 52       & 51    & -32    & 38        \\
P.A.($^{o}$)           &  150     & 112   & 65     & 57        \\
\tableline
V$_{syst}$(km.s$^{-1}$) & 6.30     & 4.89  & 6.04   &5.10       \\
V$_{100}$(km.s$^{-1}$)  &3.00      & 2.05  &2.16    &4.03       \\
$\delta_v$(km.s$^{-1}$)& 0.2      & 0.2   & 0.2    &0.2        \\
\tableline
 T$_{100}$(K)          & 15       & 15    & 15     & 30        \\
R$_{int}$(AU)          &  45      & 1     & 1      &   1       \\
R$_{out}$(AU)          & 550      & 900   & 800    &   500     \\
\tableline
\end{tabular}
\tablecomments{Following \citet{Dutrey+etal2011}. P.A. is the position angle
of the disk rotation axis, $i$ is the inclination, V$_{syst}$ is the systemic velocity
and $\delta_v$ the turbulent line-width component. The velocity laws (V$_{100}$)
are Keplerian (as shown by \citet{Pietu_etal2007} and \citet{Schaeffer_etal2009})
}
\label{tab:diskfit}
\end{table}

\begin{table*}
{\small
\caption{Derived and predicted surface densities and ratios}
\centering
\begin{tabular}{lcccccccc}
\tableline \tableline
        & \multicolumn{4}{c}{$\Sigma_{300}$ (cm$^{-2}$)}\\
        & \multicolumn{2}{c}{\hcccn} & \multicolumn{2}{c}{CCS} & \multicolumn{2}{c}{\hcccn /HCN}& \multicolumn{2}{c}{\hcccn/CN}  \\
Souce   & Derived & Predicted & Derived & Predicted &Derived & Predicted  & Derived & Predicted  \\
\tableline
LkCa\,15 & 8 $\pm$ 2$\cdot$ 10$^{11}$ & 5.2$\cdot$ 10$^{13}$ &$\leq$ 1.4$\cdot$ 10$^{12}$& 2.9$\cdot$ 10$^{11}$ & 0.075 &   0.12           & 0.014       &   0.011   \\
GO Tau   & 13 $\pm$ 2$\cdot$ 10$^{11}$ & 4.4$\cdot$ 10$^{13}$&$\leq$ 1.2$\cdot$ 10$^{12}$& 3.7$\cdot$ 10$^{11}$ &    -   &      0.08       &  - &  0.02    \\
DM Tau   & $\leq$ 3.5$\cdot$ 10$^{11}$ & 4.4$\cdot$ 10$^{13}$&$\leq$ 1.1$\cdot$ 10$^{12}$& 3.7$\cdot$ 10$^{11}$ & $\leq 0.05$ & 0.08       & $\leq$0.01  & 0.02  \\
MWC\,480   & 6$\pm$ 1$\cdot$ 10$^{11}$ & 6.4$\cdot$ 10$^{11}$&$\leq$ 0.9$\cdot$ 10$^{12}$& 3.1$\cdot$ 10$^{11}$ & 0.55  &  0.05            & 0.057  & 0.001   \\
\tableline
\end{tabular}\\
\tablecomments{Col. 2 and 4: Surface densities derived at 300 AU from the 30-m data using Diskfit (see section \ref{sec:obs}
for details). Col. 3 and 5: Predicted surface densities at 300 AU using Nautilus from CIDV best chemical model
(see section \ref{sec:dis} for details). Note that the model is the same for DM Tau and GO Tau.
Col. 6-9: Surface densities ratios at 300 AU. ``Derived ratios'' are calculated using the CN and HCN surface densities in LkCa 15, DM Tau and MWC\,480 from \citet{chapillon_etal2012}.
}
\label{tab:results-ratio}
}
\end{table*}

\section{Discussion}
\label{sec:dis}

\subsection{Chemical models of protoplanetary disks}
Chemical modeling is essential to retrieve the vertical structure of the disk from the observable column densities. 
Generic chemical models of circumstellar disks have been published by
\citet{aikawa_herbst1999}, \citet{Willacy+Langer_2000}, \citet{Aikawa_etal2002}, \citet{vanZadelhoff_etal2003}, \citet{Aikawa+Nomura_2006}, \citet{Willacy+etal_2006} and \citet{Fogel+etal_2011, Semenov_Wiebe2011, Vasyunin_etal2011}.
Each of these works make different assumption and hypothesis.
\citet{aikawa_herbst1999} consider gas phase chemistry, along with sticking and desorption, and assume a vertically isothermal disk model following the ``minimum mass solar nebula'' extended
to 800 AU. The study from \citet{Willacy+Langer_2000} use the thermal structure derived
from the \citet{Chiang+Goldreich_1997} two-layer approximation, as well as enhanced photodesorption yields
following \citet{Westley+etal_1995}. \citet{Aikawa_etal2002} use a D'Alessio disk model including vertical
temperature gradients
and self-consistently variable flaring, although dust and gas are assumed to be fully thermally coupled.
All three models use a 1+1D approximation for radiative transfer, in which the stellar UV is attenuated
along the line of sight to the star, while the ISRF impacts isotropically on the disk surface.
\citet{vanZadelhoff_etal2003} improved the UV treatment by using a 2-D radiative transfer code
to solve for the UV field inside the disk, as did \citet{Fogel+etal_2011}. Shielding by H$_2$ is treated in an approximate way, however.
\citet{Aikawa+Nomura_2006} expanded the models further by considering the effect of grain growth as \citet{Woitke_etal2009}.
\citet{Woitke_etal2009} and \citet{Fogel+etal_2011} investigated the effects of dust settling,
\citet{Fogel+etal_2011}, \citet{Willacy+etal_2006} and \citet{Walsh_etal2010,Walsh_etal2012} also include grain-surface chemistry, which may be important in such environments. 1D turbulent diffusion is included in the model from \citet{Willacy+etal_2006}, and 2D in the models from \citet{Semenov_etal2006} and \citet{Semenov_Wiebe2011}. 
\citet{Vasyunin_etal2011} have investigated dust coagulation, fragmentation, sedimentation, and turbulent stirring..

Making comparisons with our observations is difficult for several reasons.
All these chemical models are not necessarily tailored to the physical conditions relevant to the specific sources we have observed. The
chemical networks used are different from one code to another and the large uncertainties on reaction rates may explain some intrinsic
differences \citep[e.g.][]{Daranlot_etal2012}. In many cases the initial physical and chemical conditions are also different.

\subsection{The Nautilus models}
Following our previous studies, we performed a chemical modeling using Nautilus, a gas-grain chemistry model adapted for the disk physics \citep{Hersant_etal2009}. Nautilus computes the abundances of 460 gas-phase and 195 surface species as a function of time using the rate equation method \citep{1992ApJS...82..167H} and the chemical network contains 4406 gas-phase reactions and 1733
reactions involving grains, including adsorption and desorption processes and grain-surface reactions (we assume 4580 K for the energy of desorption of CCS and \hcccn). 
The gas-phase network is regularly updated (both by optimizing the reaction rates and by adding new reactions) 
according to the recommendations from the KIDA\footnote{http://kida.obs.u-bordeaux1.fr} experts.
In CIDV, we used Nautilus to perform a full chemistry modeling and only discussed the sulfur chemistry.
The initial abundances were obtained computing the chemical composition of the parent cloud as discussed in CIDV.
 We compare here our new observational
results with our best model from CIDV (model C). For the four disks, the physical conditions and elemental abundances used
are summarized in Fig.2 and Table 4 of CIDV, respectively. Model C was obtained assuming a C/O ratio of 1.2 \citep[following][]{Hincelin+etal_2011}, a sulfur and nitrogen abundance with respect to H of $8 \times 10^{-9}$ and $6.2 \times 10^{-5}$ \citep{Jenkins_2009}, respectively and a grain size of 0.1 $\mu$m with an initial cloud density of
$2 \times 10^{5}$ H.cm$^{-3}$ and a cloud age of 1 Myr. Our results are shown in Fig.\ref{fig:chemistry}.
At 1~Myr, we have an abundance of $10^{-10}$ for \hcccn\ and $4 \times 10^{-13}$ for CCS in the gas-phase and of
$2.5 \times 10^{-12}$ for \hcccn\ and $4 \times 10^{-11}$ for CCS, on grains.
The main conclusion from CIDV was that although our chemical model reproduces
the SO and CS column densities reasonably well, it fails to reproduce the upper limits
obtained on H$_2$S by at least one order of magnitude. This suggests that a fraction of Sulfur may be
depleted in mantles or refractory grains. At the high densities and low temperatures encountered around disk mid-planes,
H$_2$S may remain locked onto the grain surfaces and react to form other species preventing desorption of H$_2$S.\\

\subsection{Comparison with observations}
In protoplanetary disks,  only four N bearing molecules have been previously detected. NH$_3$ is not detected but N$_2$H$^+$  is observed with an abundance relative to H$_2$ of $\sim 10^{-12} $  \citep{dutrey_etal2007, oberg_etal2011}. CN is easy to detect \citep{Dutrey_etal1997} and likely results from a chemistry which is more
sophisticated than the simple photo-dissociation of HCN under the stellar and
ambient UV fields \citep{chapillon_etal2012}. As an example, in the DM Tau disk,
the observed surface densities of CN and HCN at 300 AU are $3.5 \cdot 10^{13}$ cm${^{-2}}$ and
$6.5 \cdot 10^{12}$ cm${^{-2}}$, respectively. HNC has been reported in DM Tau only by \citet{Dutrey_etal1997}.
\hcccn\ is the fifth nitrogen-bearing molecule detected in protoplanetary disks.
We observe this molecule in MWC\,480, GO Tau and LkCa 15.  Fig.\ref{fig:chemistry} presents the modeled and observed surface densities of \hcccn\ at 300 AU and Fig.\ref{fig:abun} presents the modeled abundances. DM Tau and GO Tau are assumed to share the same disk physical structure and therefore the chemical predictions are identical.

Fig.\ref{fig:chemistry} (left panel) shows that our non detection of CCS remains, at the 3$\sigma$ level, in very good agreement with the model C from CIDV. Furthermore, our CCS upper limits are incompatible with the other models we explored in CIDV. By increasing the elemental abundance of sulfur by a factor 10 (models A and B), we would produce ten times more CCS with respect to H$_2$.
Fig.\ref{fig:chemistry} (right panel) also shows that CCS is expected to be formed around 1.5 scale-heights in the outer disk (R$\sim 300$~AU) of both T\,Tauri and Herbig Ae stars.
\citet{Semenov_Wiebe2011} have calculated typical column densities of gaseous CCS at 300~AU to be about $10^{11}$~cm$^{-2}$ and $7\,10^{11}$~cm$^{-2}$ in the laminar and 2D-mixing models of the DM Tau disk. The CCS layer in these models is also located around 1 pressure scale height from the midplane.

 Contrary to the observed behavior of sulfur-bearing molecules (CS and H$_2$S), whose surface densities are of the same order of magnitude in both sources, \hcccn\ is only detected in the disk surrounding GO Tau. In DM Tau, the 3$\sigma$ upper limit is a factor 4 lower than the detection in GO Tau. Some intrinsic differences in the physical/chemical properties have to be explored to understand the origin of such a difference.

\subsection{Results}
\begin{figure}
\centering
\includegraphics[width=8.5cm]{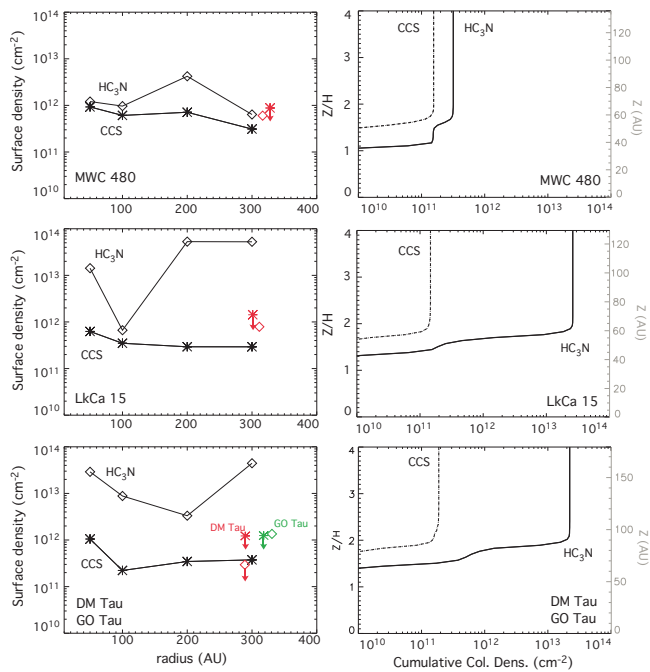}
\caption{ Predicted column densities for \hcccn and CCS using Nautilus. Left panel: observed and modeled surface densities. Right panel: Cumulative column densities (from mid-plane to atmosphere) in function of $Z/H$ (left axis) and Z (right axis) at 300\,AU.
Top panel: MWC\,480, medium panel: LkCa15, Bottom panel: GO Tau and DM Tau. The (cumulative) column densities (right panel),
whose asymptotic values are half of the surface densities (left panel).}
\label{fig:chemistry}
\end{figure}

For the T\,Tauri disks, our best chemical model predicts \hcccn\ surface densities which are typically two orders of magnitude larger than the observed one. Our chemical model show that \hcccn\  is found in a layer around 1-2 scale height in the T\,Tauri disks (see Fig.\ref{fig:chemistry}). In this
layer,  the density is of the order of $\sim 3\times 10^{6}$cm$^{-3}$, i.e. comparable or slightly above the expected critical densities of the transitions \citep{Wernli_etal2007}, especially for J=16-15 transition. As we assumed LTE, this may affect our determination of the surface
densities by a factor of at most a few, but not as large as $\sim$ 100.

The chemical model however appears in good agreement for the disk surrounding the Herbig Ae star MWC\,480 in which the predicted abundance is much smaller than in the other sources. We checked that the difference is due to the UV flux, significantly
stronger in the case of MWC\,480, which photo-dissociates \hcccn. The \hcccn\ abundance appears better reproduced in the presence of a strong UV field. Enhancing the UV penetration by changing vertically the dust grain properties (grain growth and vertical settling) may decrease the predicted \hcccn\ column densities even in sources with low or medium UV fluxes. The stronger UV flux also explains that most of the \hcccn\ is formed at an altitude which is lower in the Herbig Ae disk ($Z/H \simeq 1$) than in T\,Tauri disks ($Z/H \simeq 1.5$, see Fig.\ref{fig:chemistry}, right panel). As are result, the observed \hcccn\ transitions should be more easily thermalized in the Herbig Ae disk.\\

\begin{figure}
\centering
\includegraphics[width=8.5cm]{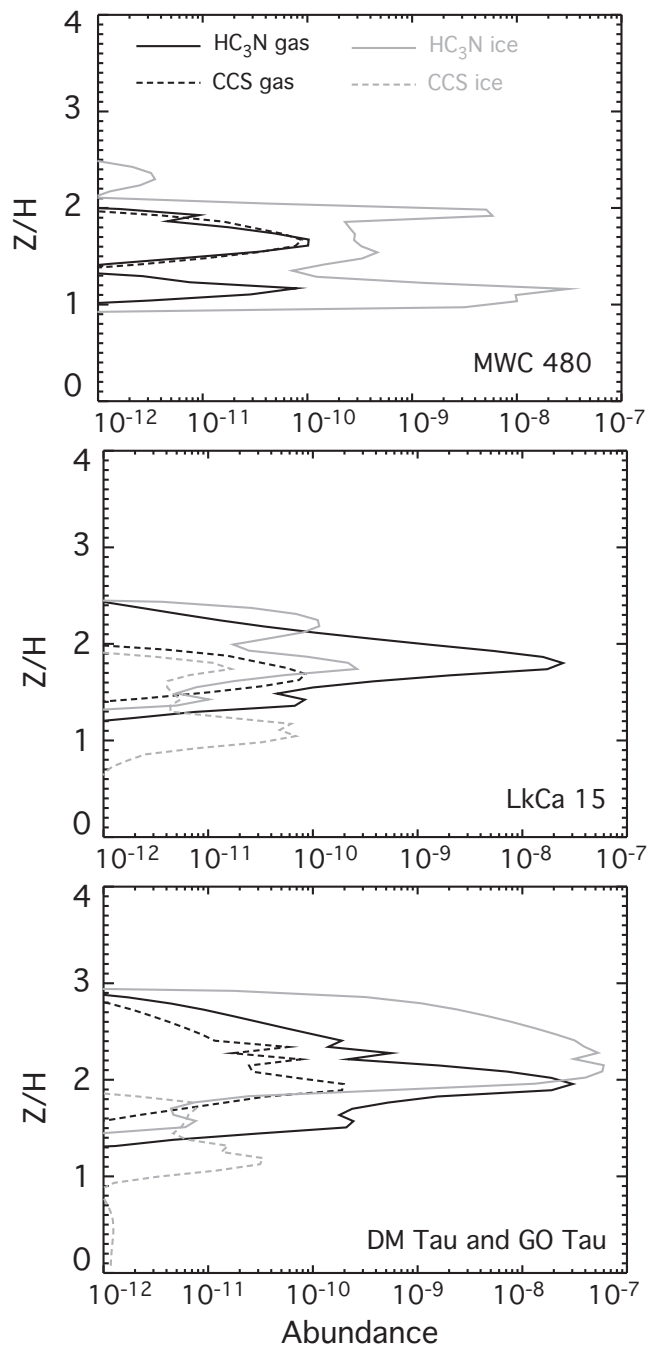}
\caption{Abundances of \hcccn\ (solid line) and CCS (dashed line) in the vertical direction at 300 AU in the four
protoplanetary disks. The black and grey lines represents the abundances in the gas-phase and in the ices.}
\label{fig:abun}
\end{figure}

Investigating the vertical abundance variation is also very intersting. Figure \ref{fig:abun} shows the predicted vertical abundances of \hcccn\ and CCS, in the gas-phase and at the surface of the grains at 300 AU in the four disks. These two molecules are present in a layer between 1 and 3 scale heights in DM Tau and at lower height in the two other disks where the stellar UV flux is stronger (410, 2550, 8500 $\chi_0$ for DM Tau, LkCa 15 and MWC 480 respectively, see CIDV). In all models, CCS is formed in the gas-phase only and can only be destroyed on grain surfaces, after being accreted. In DM Tau, the surface abundance of CCS is smaller than in LkCa 15 because more complex S-bearing species such as C$_3$S are formed. In LkCa 15, the formation of such molecules is prevented by the stronger UV field. In MWC 480, the CCS abundance in the ices is even smaller than $10^{-12}$. The gas phase abundance of \hcccn\ for DM Tau is similar to the one in LkCa 15, whereas the abundance in the grain surface is much larger. The smaller surface abundance of \hcccn\ in LkCa 15 is due to direct photo-dissociations by UV photons. The situation in MWC 480 is very different, with a very large abundance of \hcccn\  in the ices and a very small one in the gas. Although the UV field is stronger than in the two other sources, the dust temperature is larger and allows the diffusion of larger radicals on the surfaces. This enhanced diffusion produces more \hcccn\ on grains.\\

In Table \ref{tab:results-ratio}, we compare our results with those from \citet{chapillon_etal2012}.
If our prediction of the \hcccn\ column density in MWC\,480 is in good agreement with the observation, it fails to reproduce the \hcccn/CN and \hcccn/HCN ratios. For the T Tauri disks, the ratios are in roughly good agreement with the observations while the predicted absolute values are off by two orders of magnitude. This is not surprising since this model was aimed at reproducing
sulfur-bearing molecule chemistry. Using a lower elemental abundance of nitrogen would produce less N bearing molecules, \citep[see][]{Semenov_Wiebe2011}. Moreover, some neutral-neutral reactions involving atomic nitrogen are currently being experimentally studied \citep{Bergeat_etal2009, Daranlot_etal2011, Daranlot_etal2012}. As soon as these new experimental rates are known, a more detailed study of nitrogen-chemistry in disks will be performed \citep{Wakelam_etal2012}. We also note that the observed CN/HCN ratio is of the order of 5 in the T Tauri stars disks, and of 10 in the disks of Herbig stars.

The observed \hcccn/HCN ratio in the three disks is very similar to that observed in the cold molecular core L134N \citep[HC$_3$N/HCN $\sim$ 0.07,][]{Dickens_etal2000} and still reasonably close to the cometary value within a factor $\sim 2$ \citep[in Hale-Bopp, HC$_3$N is observed with a ratio HC$_3$N/HCN $\sim$ 0.1,][]{Bockelee-morvan_etal2000}. However, the observed \hcccn/CN ratio of both Herbig Ae and T\,Tauri disks is far from the value observed in L134N \citep[HC$_3$N/CN $\sim$ 1,][]{Dickens_etal2000}, suggesting that the later ratio is more affected by photo-dissociation.\\

These observations can be used to place the detectability of complex molecules in perspective with ALMA. Our typical 3$\sigma$ limit is of order 50 mJy.km.s$^{-1}$ for a single line (Table \ref{tab:fluxes}), but we gain $\sqrt{3}$ by the multiplex factor due to wide frequency coverage. By comparison, ALMA reaches 1 mJy in 1 hour of integration time for 1 km.s$^{-1}$ resolution at 90 GHz, i.e.1$\sigma$ flux sensitivity of 1.4 mJy.km.s$^{-1}$. ALMA should greatly improve the detectability of such molecules, but imaging at high ($\leq 0.5''$) resolution would require many (more than 10) hours.
\section{Summary}

Using the 30-m IRAM telescope, we report the first detection of \hcccn\ in three disks surrounding young low-mass stars and present
\emph{deep} upper limits on the CCS molecule. Our results allow us to conclude:
\begin{itemize}
\item The CCS upper limit confirms our previous modeling of sulfur-bearing molecule which favors 
 a relatively low elemental abundance of sulfur ($8 \cdot 10 ^{-9}$/H).
\item The observed column densities of \hcccn\ are typically two orders to magnitude lower than the predictions.
\item A strong UV flux decreases the abundance of \hcccn\
 suggesting that a better UV penetration either due to significant grain growth or dust settling may significantly affect the \hcccn\ surface density, even when a moderate UV flux impacts the disk.
\item More generally, our observations confirms that complex molecules remain difficult to detect in disks.
\end{itemize}

\begin{acknowledgements}
We acknowledge all the 30-m IRAM staff for their help during the observations. This research was partially supported by PCMI, the French national program for the Physics and Chemistry of the Interstellar Medium.
D.S. acknowledges support by the Deutsche Forschungsgemeinschaft through SPP 1385: "The first ten million years of the solar system—a planetary materials approach" (SE 1962/1-2).
\end{acknowledgements}

%%%%%%%%%%%%%%%%%%%
\end{document}